\documentclass{elsarticle}
\usepackage{graphicx}
\usepackage{psfrag}
\usepackage{makeidx}  
\usepackage{multirow}
\usepackage{psfrag}
\usepackage{caption}
\usepackage{amsmath}
\usepackage{wrapfig}

\begin{document}

\title{ Designing a High Performance Parallel Personal Cluster\footnotemark[1]}

\author[bg]{K.G.~Kapanova$^*$\footnotemark[2]} 
\author[bg]{J.M.~Sellier\footnotemark[2]}
\address[bg]{IICT, Bulgarian Academy of Sciences, Acad. G.~Bonchev str. 25A, 1113 Sofia, Bulgaria\\$^*$\texttt{kapanova@parallel.bas.bg,\\kkapanova@gmail.com}}

\begin{abstract}
Today, many scientific and engineering areas require high performance computing to perform computationally intensive experiments.
For example, many advances in transport phenomena, thermodynamics, material properties, computational chemistry and physics are
possible only because of the availability of such large scale computing infrastructures. Yet many challenges are still open.
The cost of energy consumption, cooling, competition for resources have been some of the reasons why the scientific and
engineering communities are turning their interests to the possibility of implementing energy-efficient servers utilizing
low-power CPUs for computing-intensive tasks. In this paper we introduce a novel approach, which was recently presented at
Linux Conference Europe 2015, based on the Beowulf concept and utilizing single board computers (SBC). We present a low-energy
consumption architecture capable to tackle heavily demanding scientific computational problems. Additionally, our goal
is to provide a low cost personal solution for scientists and engineers. In order to evaluate the performance of the
proposed architecture we ran several standard benchmarking tests. Furthermore, we assess the reliability of the machine in
real life situations by performing two benchmark tools involving practical TCAD for physicist and engineers in the semiconductor industry.

\footnotetext[1]{\texttt{This work has been submitted to a journal on 14 November 2015}}
\footnotetext[2]{\texttt{These two authors contributed equally to this work}}

\end{abstract}


\maketitle


\section{Introduction}
\label{Introduction}
Nowadays, many different areas of science and engineering require high performance computing (HPC) to perform
data and computationally intensive experiments \cite{Prabu}. Advances in transport phenomena, thermodynamics,
material properties, machine learning, chemistry, and physics are possible only because of large        scale
computing infrastructures. Yet, over the last decade the definition of high performance computing      changed
drastically. The introduction of computing systems made from many inexpensive processors became an alternative
to the large and expensive supercomputers. In $1994$ the introduction of a practical way to link many personal
computers in a computing cluster represented cost-effective and highly performance model \cite{Beowulf}. Today,
the Beowulf concept represents a recognized subcategory within the high-performance computing community.

This vision, although implemented in certain way, is not entirely realized and many challenges are still open.
For instance, the cost of energy consumption of a single server during its life cycle can surpass         its
purchasing cost \cite{Barroso}. Furthermore, at least $33\%$ of the energy expenses in large-scale   computing
centers are dedicated to cooling \cite{Greenberg}. Need for energy preservation have forced many HPC   centers
to relocate near power stations or renewable energy location and to utilize natural/green resources like   sea
water or outside air in naturally cold areas in the world. Idle resources account for a portion of the  energy
waste in many centers, and therefore load balancing and workload scheduling are of increased interest in  high
performance computing. Furthermore, depending on the specific computational tasks, it is more cost   effective
to run more nodes at slower speed. In other cases exploring parallelism on as many processors as      possible
provides a more energy efficient strategy \cite{Freeh}, \cite{Langen}. As a consequence from the     different
computational demands a competition for resources emerges. Subsequently, the scientific and        engineering
communities are turning their interests to the possibility of implementing energy-efficient servers  utilizing
low-power CPUs for computing-intensive tasks.

Taking under consideration the above factors, in this paper we present a novel approach to          scientific
computation based on the Beowulf concept, utilizing single board computers (SBC). Our goal is two         fold.
Firstly, we want to show a low-energy consumption architecture capable to tackle  heavily demanding scientific
computational problems. Second, we want to provide a low cost personal solution for scientists and engineers.
Recently this architecture has been presented at Linux Conference Europe 2015 \cite{LinuxCon}.

The structure of this paper is as follows. In the next section we describe the specific hardware we have used
to build our cluster. Then we outline the software packages the server runs on. We provide several  benchmark
tests to assess the performance of our suggested architecture. Furthermore, we carry out two benchmark  tools
involving practical TCAD for physicist and engineers in the semiconductor industry in order to assess     the
reliability of this machine in real life situations. In more detail we run two well-known GNU      packages -
Archimedes \cite{Archimedes} and nano-archimedes \cite{nanoarchimedes} which respectively deal with classical
(CMOS technology) and quantum electron transport (post-CMOS technology). We conclude the paper with  possible
future directions.

\section{Developing the High Performance Parallel Personal Cluster}
\label{development}

The single board computer is a complete computer architecture on a single circuit board. Compared to its low
cost, the board provides relatively  high computational power, low power consumption and small space required
for storage. In the last few years over $40$ different SBC boards of varying architectural and computational
capabilities have been introduced \cite{Boards}. Consequently, several teams have considered and implemented
SBC boards as Beowulf clusters \cite{Parallela}, \cite{MontBlanc}. While interesting projects are concentrating
towards large centers and research institutions, others have demonstrated a Beowulf cluster consisting of
$64$ Raspberry Pi boards was described in \cite{Iridis}. 

Several factors need to be taken into consideration when developing such cluster -  computing power, memory
size and its speed, energy consumption levels, and in our specific case mobility of the cluster.
Despite the success of the Iridis-Pi project, it is rather expensive since it had to use so many Rasbperry Pi
boards to reach computational power and memory availability. Furthermore, expansion of the number of boards
leads to an increase of energy consumption and a need for cooling implementation. Those reasons hinder    the
mobility of the cluster. Finally, while they provided a proof of concept, most projects focus on proving the
concept \cite{Boise}, applying it as a web server \cite{webserver} or for testing distributed       software
\cite{distributed} or distributed updates \cite{resin}.

The following subsections describe the most important hardware parameters of the Radxa Rock board, the cluster
architecture and the software we have ran on it. The reader should keep in mind that the type of      boards,
architecture and software are not a necessary limitations and other specifications can be implemented as well.

\subsection{Hardware Specifications}
\label{cluster}
In this section, we present the most relevant hardware parameters of the Radxa rock board (in the context
of our project). The selection of this specific board depended on several primary factors: computing power,
memory size and speed, communication speed. The size of the boards and the case we have built are also
considered since heat dissipation is an essential factor for the necessary cooling strategies and therefore
operational costs of the machine. We have concentrated on developing an architecture that is mobile
and easy for transportation but at the same time able to perform meaningful          science/engineering
computational tasks. In terms of usability, determining the software needs and specific packages to be
employed has been taken under consideration.  
\bigskip

\begin{figure}[h!]
\centering
\begin{minipage}{0.5\textwidth}
\begin{tabular}{c}
\includegraphics[width=1.0\textwidth]{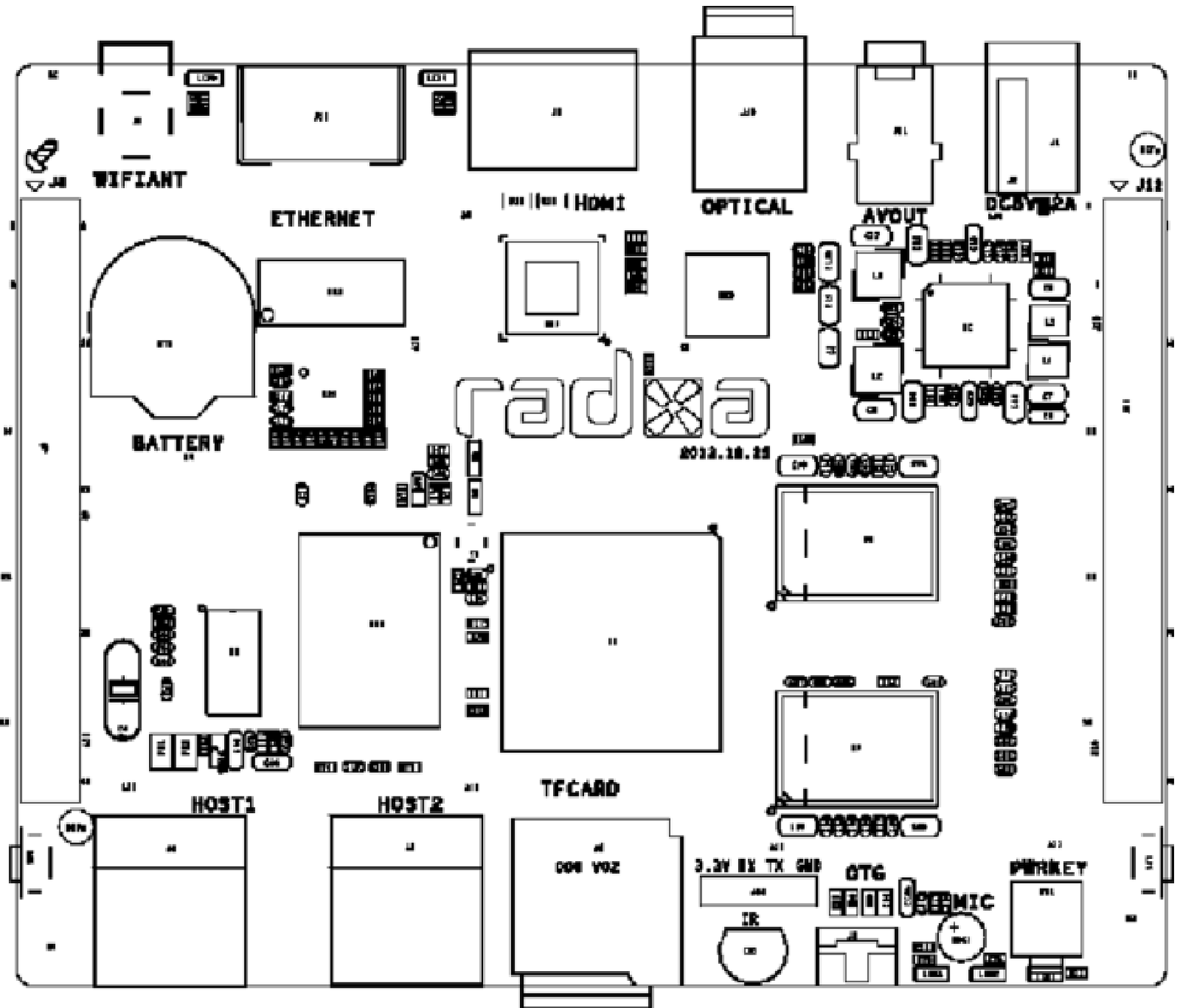}
\end{tabular}
\end{minipage}
\caption{Schematic from the top of the component position for Radxa Rock \cite{Radxa}. The Radxa Rock is an
Open Source Hardware design, with all the necessary design documentation  being licensed through the
\textbf{Creative Commons Attribution 4.0 International License}.}
\label{fig:radxacomp}
\end{figure}

Our High Performance Parallel Personal Cluster ($HP^3$) consists of $4$ Radxa Rock Pro        nodes
\cite{Radxa} (see fig.~\ref{fig:radxacomp}), with a plug-and-play model (specifically developed for our purpose)
for the addition of more boards when we require more computational power. The Radxa board itself has a Rockchip $RK3188$ $ARM Cortex-A9$
CPU architecture with $4$ cores running at $1.6GHz$. The board is equipped with $2GB$ of DDR3 RAM (at $800MHz$),
which has considerably more bandwidth compared to, for instance, Raspberry Pi or the BeagleBone boards
\cite{Raspspec}, \cite{Beagle}. Currently, the
system-on-chip boards are not designed to extend the physical RAM, but since we are using   distributed
memory provided by all boards, this disadvantage is mitigated to a certain level. The non-volatile storage on
the board is a $8GB$ of NAND Flash memory. Currently, this space is enough to install the      required
operating system and all the necessary software in order to perform highly sophisticated        quantum
computations (see section \ref{compute}). Furthermore, the storage capacity can be extended through the available
microSD card slot, which can be used to increase the virtual memory of the cluster or to add more storage
capacity. In the current work, we use a generic $16GB$ SDHC flash memory with a minimum write speed of $10MB/s$.
We have used part of the additional storage as a virtual memory and the rest as a storage space.    The
purpose of the latter is to reduce the size of the software that is needed by each node and     instead
provide an option to compile it dynamically. Therefore, we have implemented a Samba share to    provide
access to all client nodes in the system. The decision to choose Samba is \textit{by no means} a limiting
factor and one might instead use NFS or rsync in similar manner. In this way,s we provide access to the software
and data to all available nodes. Additionally, we can implement upgrades and changes without increase of workload
time or wasting memory in every board. If specific libraries (such as MPI) are needed for certain computations,
they are installed on the external drive, with hard links from every node pointing to their storage location.

The board provides also two USB $2.0$ ports, a $10/100Mbits$ Ethernet adapter,  built-in \textit{Wifi}
module at $150Mbp$ with support for the $802.11b/g/n$ protocol and a $Bluetooth 4.0$ module. The graphical
processing unit is Mali-$400-mp4$ running at $533Mhz$, with one HDMI output. 

\subsection{Connections}
\label{Connections}
\begin{figure}[h!]
\centering
\begin{minipage}{0.8\textwidth}
\begin{tabular}{c}
\includegraphics[width=0.8\textwidth]{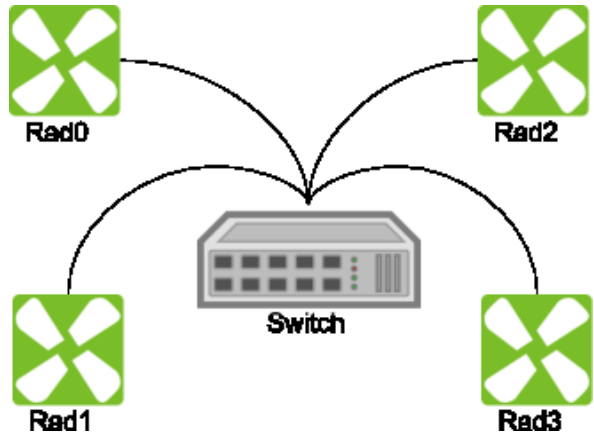}
\end{tabular}
\end{minipage}
\caption{Star network topology for the current implementation of the cluster.}
\label{fig:star}
\end{figure}

In this paper, we use the standard $100 Mbps$ network interface as the inter-node communication for the
$HP^3$ cluster. The communication is accomplished through an $8$-port Ethernet $100Mbp$ switch in    a
typical star configuration (see fig.~\ref{fig:star}). Accordingly, every board is connected directly to
the switch. While in most Beowulf architectures, one node serves as the master one, which is responsible
for user management and maintaining the correct software on the specified nodes, it does not always contribute
to the specific computational tasks. In our case, we do define a master node, and we are also utilizing
it for computational purposes. Contrary to other SBC boards, the Media Access Control(MAC)  address for
every node is not unique for every board as assigned by the manufacturer. Instead, all boards come with
the same MAC address which is detrimental to the communication of the cluster. Therefore, during    the
setup of the cluster, one has to assign unique MAC address to the Network interface controller of every
board.

\subsection{Power, Cooling and Case}
\label{PCC}
Every board has its own USB OTG power cable with recommended power supply of $5V/2A$. For our cluster
implementation, we have build a custom power supply unit using an ATX power supply. The power unit is
also capable of providing power for the cooling fans if such are needed. 

Although many cluster implementations rely on powerful cooling systems needing a considerable amount of power,
the reader should note that in our specific case for the $HP^3$ we are currently not using any cooling method.
Instead, we have build a custom case, which provides \textit{natural cooling} flow for te boards, switch and 
power supply. In case cooling is needed, the case is built to house to general purpose cooling fans (each $80$mm)
underneath, which connect to the custom power supply unit.

\begin{figure}
\centering
\begin{minipage}{1.0\textwidth}
\begin{tabular}{c}
\includegraphics[width=5cm, height=7cm]{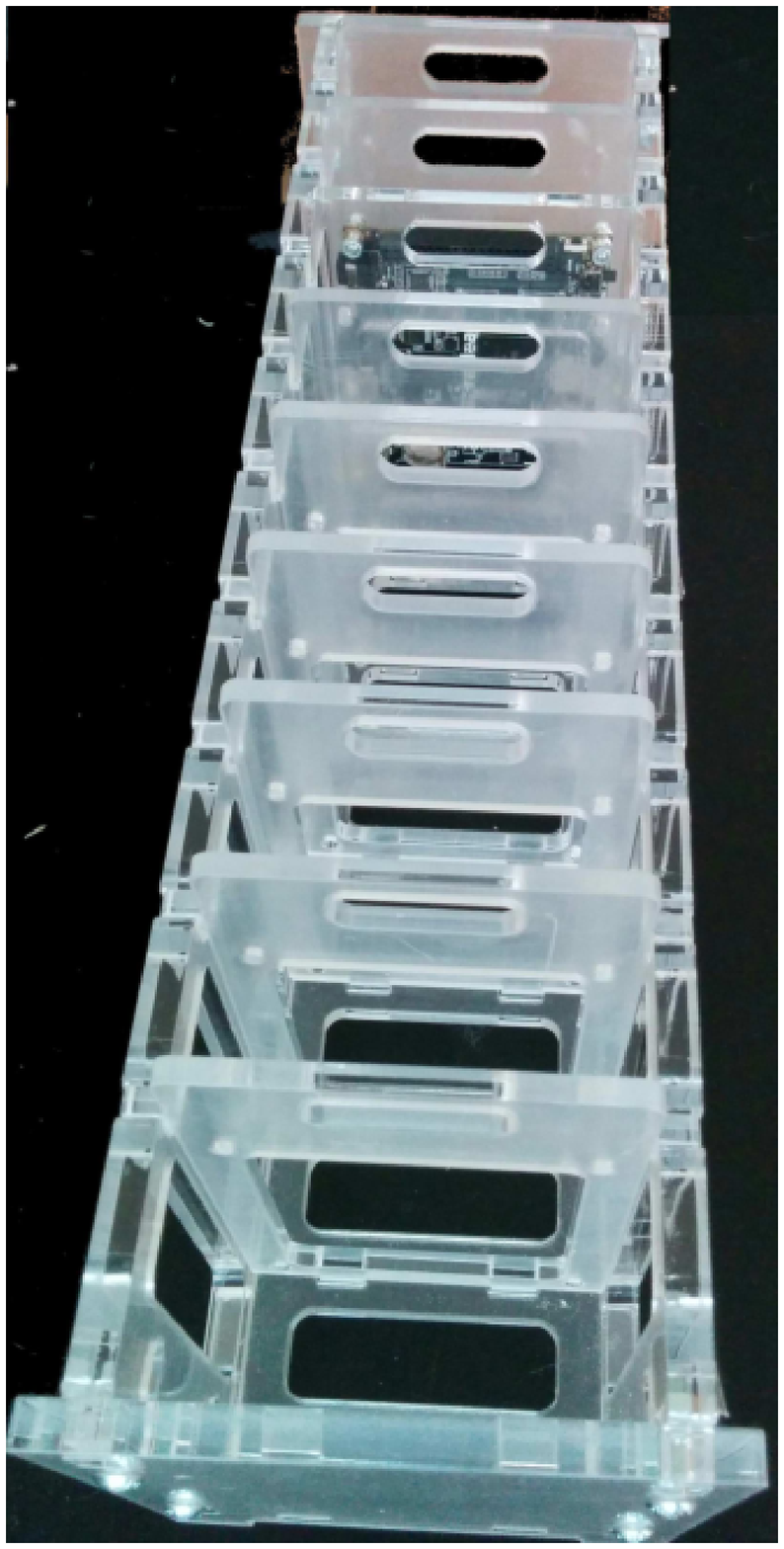}
\includegraphics[width=5cm, height=7cm]{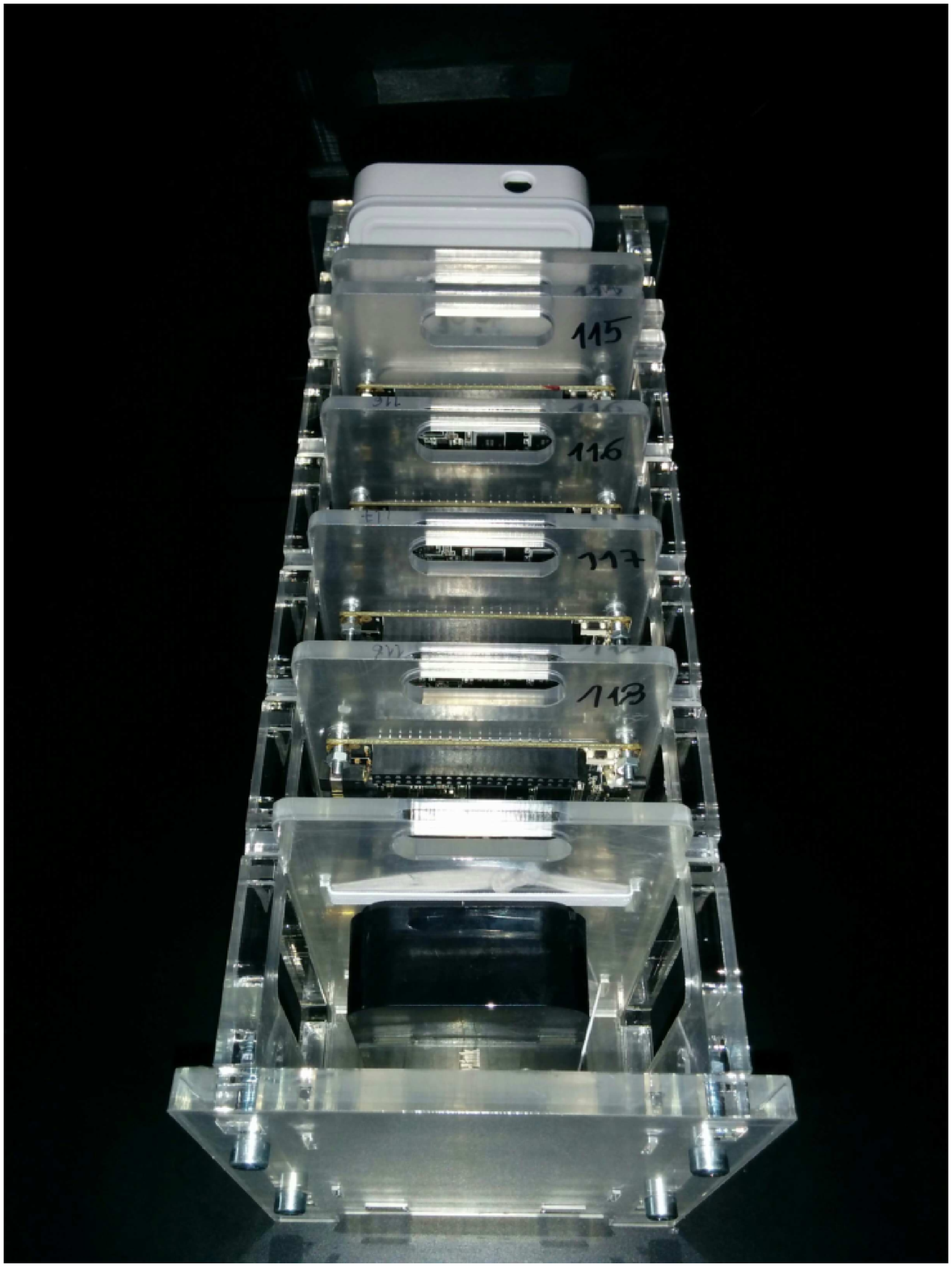}\\
\includegraphics[width=5cm, height=7cm]{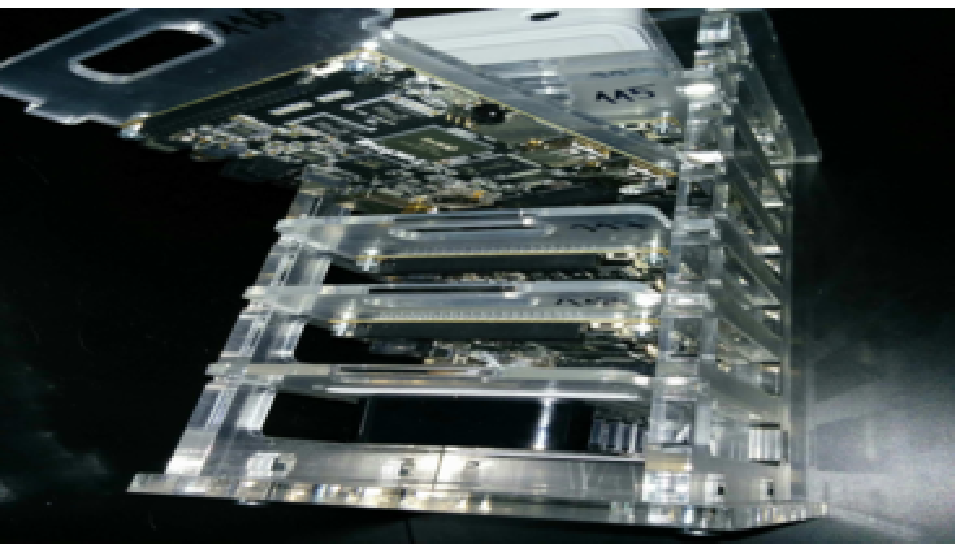}
\includegraphics[width=5cm, height=7cm]{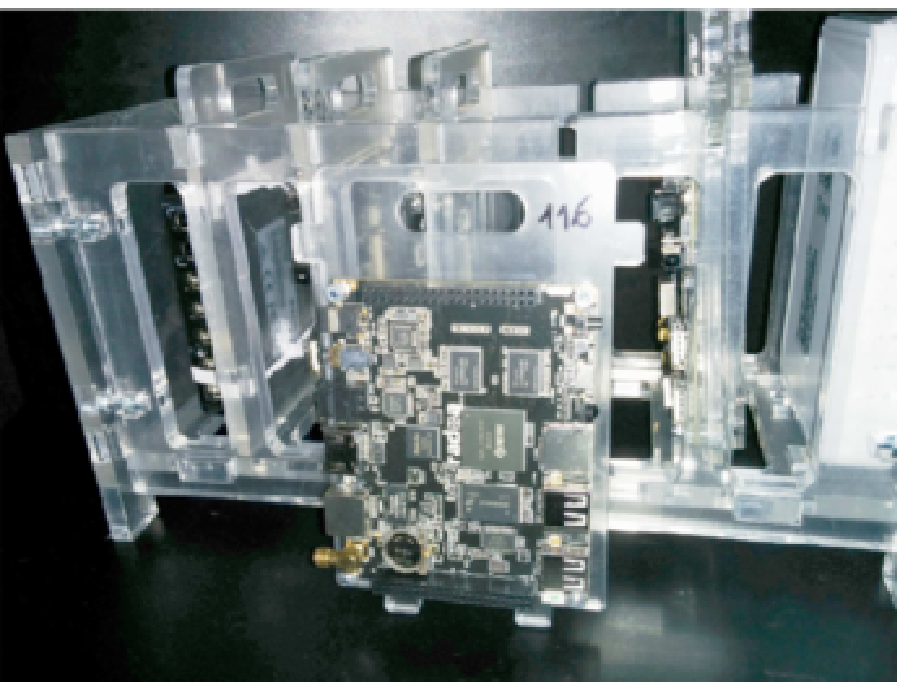}
\end{tabular}
\end{minipage}
\caption{Pictures representing our particular implementation suggested in this paper. In particular it is possible to observe
the plug-and-play technology in the project.}
\label{fig:machine}
\end{figure}

\subsection{Service Software}
\label{Software} 
Every Radxa board is shipped with a pre-configured Android and Gnu/Linux Distribution of Ubuntu.  Such
configuration was not suited for our specific scientific computation needs. Respectively, we have removed
everything on the NAND flash memory and installed an optimized and custom deployed GNU/Linux distribution - Linaro.
As an example, we have stripped the kernel support for multimedia and interactive devices. Additionally,
we have added support for the system to mount through the file system. Further customization required
the installations of several packages utilized for the benchmarking tests and application testing.

\section{Cluster Performance Evaluation}
\label{performance}
In this section, we present some results that were obtained by standard evaluation tools,
as well as in real-life scientific computations. The reader should note that all the performance
analysis was done on the configuration described in section \ref{development}. We have run some computational
benchmarks results obtained by running NAS Parallel Benchmarks (NPB) tools on
the $HP^3$ cluster designed and supported by Numerical Aerodynamic Simulation (NAS) Program at NASA Ames
Research Center. The NPB tools are implemented for studying parallel processor systems. The tool set consists
of several different benchmark problems, each focusing on important aspects of parallel computing. These benchmarks
are based either on Fortran-90 (including Fortran-77 as a subset) or C \cite{NASA}.

The first test we have ran on the $HP^3$ cluster consists of a kernel benchmark - the Embarrassingly Parallel (EP)
benchmark. The test provides an estimate of the upper achievable limits for floating point performance through
generation of pairs of Gaussian random deviates. This particular problem is typical for many Monte Carlo simulation applications.
We implement the EP test with a class B problem size. Since we have $4$ boards, each consisting of $4$ cores,
we ran several performance test with $1, 4, 8, 12, 16$ cores. When all four Radxa nodes are utilized, the EP.B
test is running at $52.19$ Mop/s, while only at $4$, the result is $13.71$ Mop/s. The total time needed for
the test when only $4$ cores are utilized is $156.64$ seconds. On the other hand, if only one core is used
the total performance is $5.57$ Mop/s. While when all cores are exploited, the time decrease 
of $73.74\%$ to $41.14$ seconds (see table \ref{table-ep}). The reader should note that this type of tests
are influenced by the communication channel of the cluster, in our case $100Mbp$ connection. Nevertheless, the
cluster performs as anticipated on a standard test case. 
\bigskip

\begin{tabular*}{\textwidth}{@{}@{\extracolsep{\fill}}rrr}
\hline
\cline{1-3}
		\multicolumn{1}{r}{Cores}               
                 &\multicolumn{1}{r}{Time in Second}
                 & \multicolumn{1}{r}{Rate Mop/s} \\
\hline
$1$ & $385.89$ & $5.57$  \\
$4$ & $156.64$ & $13.71$  \\
$8$ & $78.26$ & $27.44$  \\
$12$ & $54.67$ & $29.28$  \\
$16$ & $41.14$ & $52.19$  \\
\hline
\multicolumn{3}{@{}p{70mm}}{EP test tool for Problem size class B}
\end{tabular*}
\label{table-ep}

\bigskip
The second test used from the NPB tool set is the FT Benchmark. It contains the computational kernel of a
$3-D$ fast Fourier Transform (FFT)based spectral method. FT performs three one-dimensional ($1-D$) FFT’s,
one for each dimension. Performing this test requires the number of cores to be a power of two, thus
the available tests are for $4, 8$ cores (see table \ref{table-ft} for more details). The decrease of performance
for this test when more than $4$ cores are utilized is not a surprising effect. For instance, the performance
decreases due to all-to-all communication limitations due to bandwidth bottleneck.
\bigskip

\begin{tabular*}{\textwidth}{@{}@{\extracolsep{\fill}}rrr}
\hline
\cline{1-3}
                 \multicolumn{1}{r}{Cores}
                 &\multicolumn{1}{r}{Time in Second}
                 & \multicolumn{1}{r}{Rate Mop/s} \\
\hline
$4$ & $36.30$ & $196.62$  \\
$8$ & $45.59$ & $156.52$  \\
\hline
\multicolumn{3}{@{}p{70mm}}{FT test tool for Problem Class A}
\end{tabular*}
\label{table-ft}

\subsection{Energy Consumption}
\label{energy}
The power consumption was measured along with all the necessary communication devices and therefore, our
results can be higher than the provided by manufacturers.  The consumption was measured during the EP.B
test and included energy load in idle mode for the, only when one core was utilized, and when four cores
were loaded fully. When the board is idle, the power consumption is $3.02W$, while when one core is used,
the consumption is slightly bigger - $3.18W$, with over $3.70W$ when all cores of one board are used. 

\subsection{Computational Capabilities}
\label{compute}

In order to clearly prove that our suggested architecture can achieve real life calculations
useful for scientific investigations and engineering tasks, we decided to run the two GNU packages
Archimedes \cite{Archimedes} and nano-archimedes \cite{nanoarchimedes}, which simulate classical (CMOS technology)       and
quantum electron transport (post-CMOS technology) respectively.

\begin{figure}[h!]
\centering
\begin{minipage}{1.0\textwidth}
\begin{tabular}{c}
\includegraphics[width=0.55\textwidth]{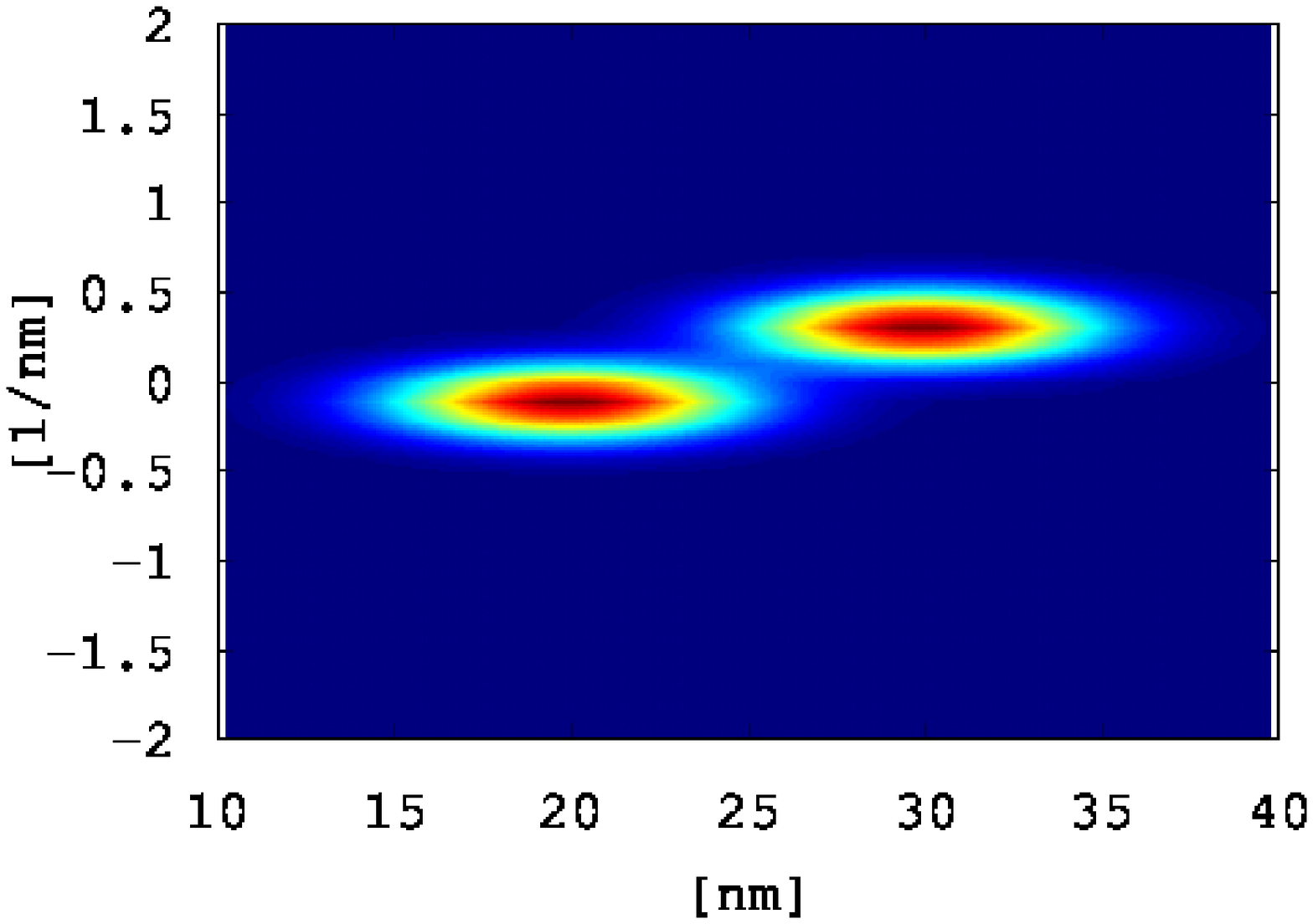}
\includegraphics[width=0.55\textwidth]{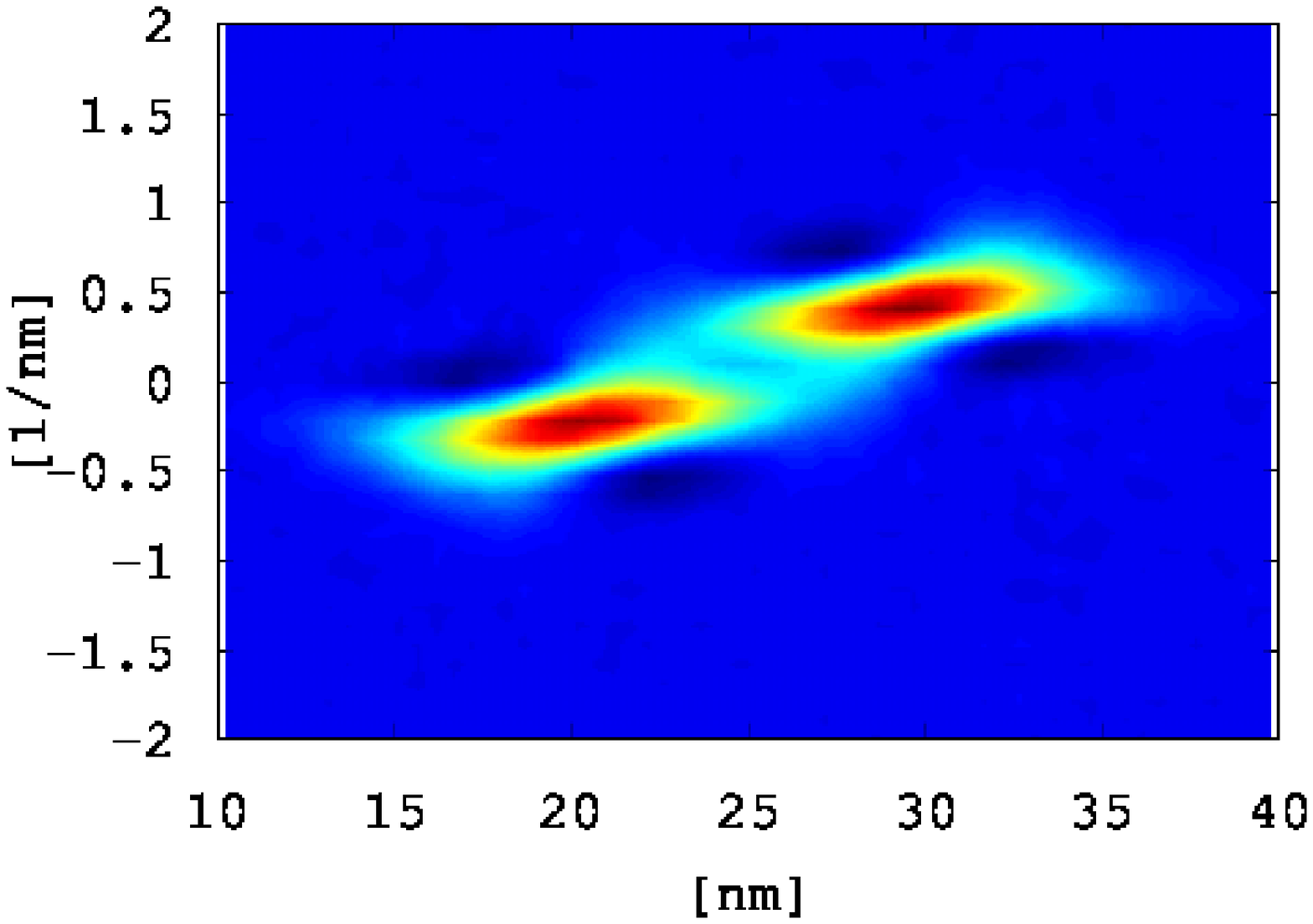}
\\
\includegraphics[width=0.55\textwidth]{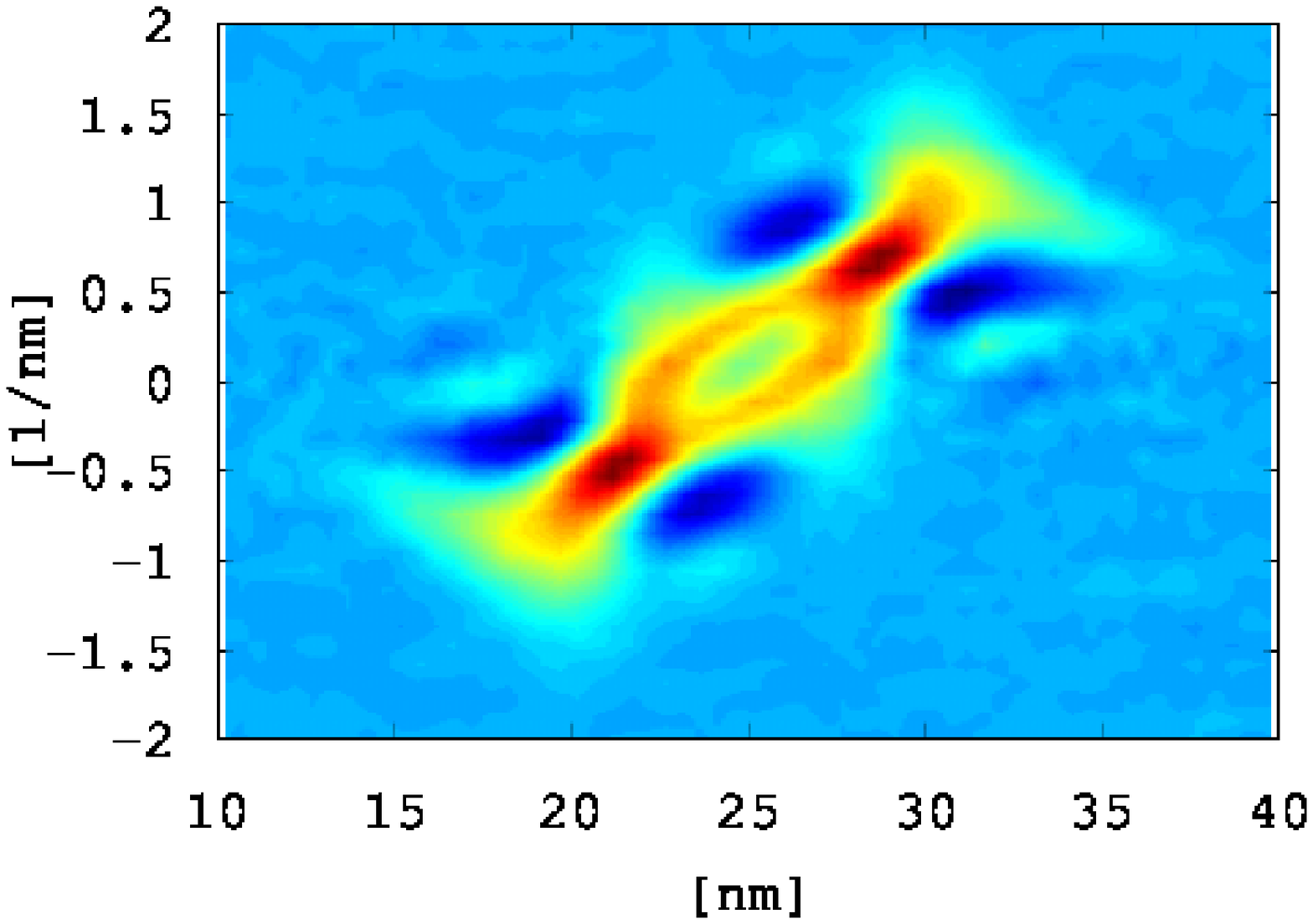}
\includegraphics[width=0.55\textwidth]{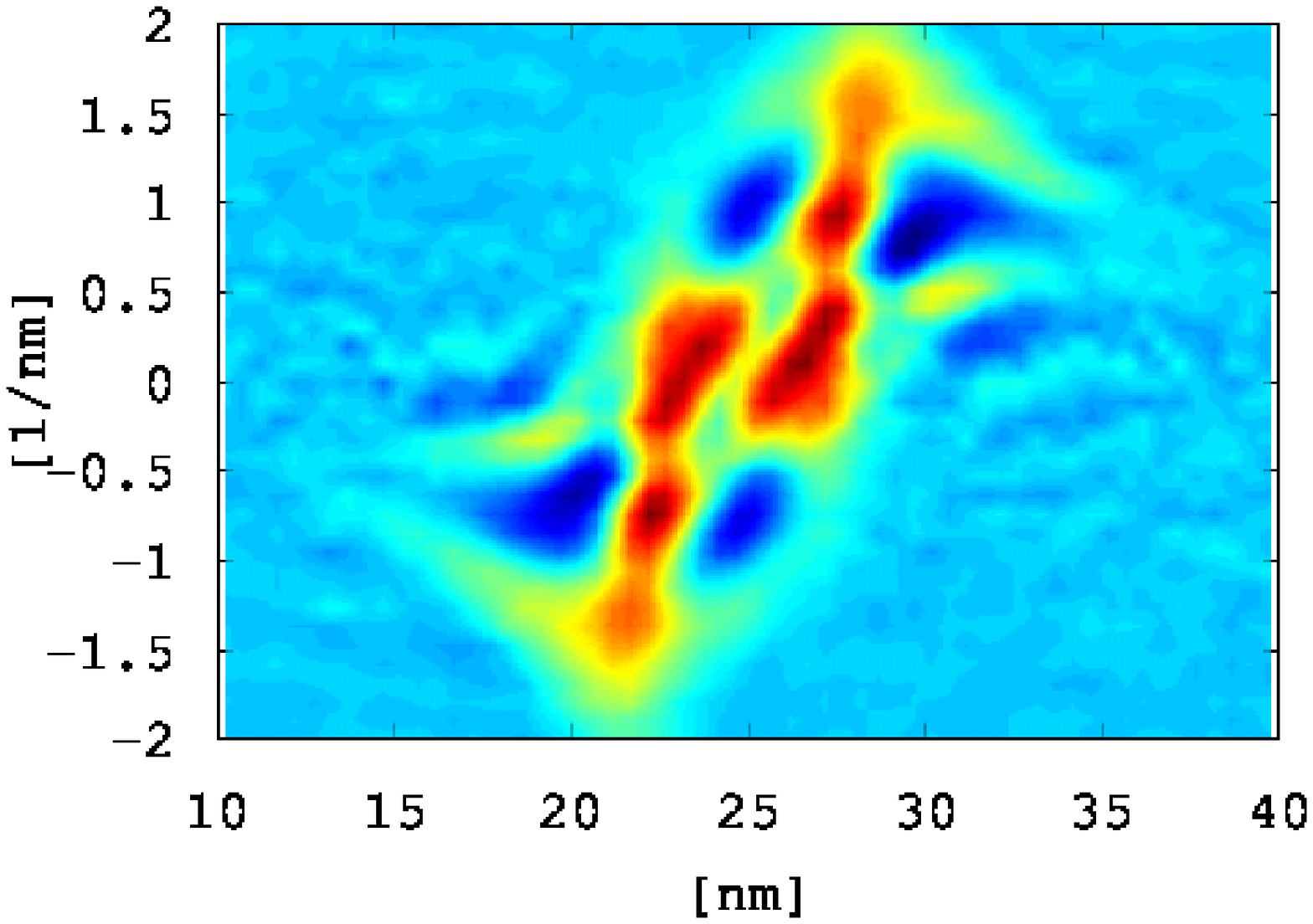}
\end{tabular}
\end{minipage}
\caption{Time-dependent evolution of the reduced one-particle Wigner function at time 0fs
(top, left) and 1fs (top right), 2.5fs (bottom left) and 3.5fs (bottom right).
The formation of a Fermi hole (also known as an exchange–correlation hole), due to the Pauli
exclusion principle, is clearly visible at time 2.5fs (bottom, left). Eventually the hole
disappears (3.5fs) as the system evolves (bottom, right). The x-and y-axes refer to position and
momentum respectively \cite{SellierDimov}.}
\label{fig:Fermihole}
\end{figure}

In particular, the {\sl{Archimedes}} package simulates CMOS semiconductor devices at a
semi-classical level. It is based on the Boltzmann Monte Carlo approach \cite{Jacoboni},
where an ensemble of classical particles miming electrons, subject to phonon scattering,
are coupled to the Poisson equation describing an external and self-generated electrostatic
potential. These particles are independent, which drastically reduces the amount of
communication between nodes. Furthermore, the code is capable of including some (first-order)
quantum effects occurring in submicron devices by means of the Bohm effective potential.
On the other hand, we also tested the cluster by running {\sl{nano-archimedes}} which aim
is the full quantum simulation of electron transport in post-CMOS technology related devices.
It is based on the (signed particle)Wigner Monte Carlo \cite{PhysicsReport} which corresponds
to a numerical discretization of the novel signed particle
formulation of quantum mechanics \cite{Formalism}. Even in this particular case, physical
systems are approached by means of Newtonian (signed) particles which are independent from
each other. In spite of the simplicity of this recent formulation, the formalism is
actually able to simulate in a {\sl{first principle}} and {\sl{time-dependent}} fashion many-body systems.
We, therefore, decided to simulate a peculiar quantum phenomenon known as the formation of
a {\sl{Fermi hole}} (also known as an exchange-correlation hole), see Fig.\ref{fig:Fermihole}.

\section{Conclusion and Perspectives}
\label{Conclusions}

In this work, we suggested a novel cluster architecture based on Systems on a Chip, which was recently presented
at the Linux Conference 2015 \cite{LinuxCon}.
We have successfully validated our practical hardware implementation over a set of standard benchmarking test (NASA, NPB \cite{NASA}).
Furthermore, in spite of the common believe of the scientific computational communities, we have been able to show
that this machine can actually perform {\sl{real life}} related simulations, in particular in the field
of CMOS and post-CMOS device design.
The reader should note that our current implementation consists of a {\sl{homogeneous}} structure
based on Radxa Rock boards only, connected by means of Ethernet interface. It is clear that these practical
decisions do not represent a restriction. As a matter of fact, we plan to design and implement a {\sl{heterogeneous}}
cluster, which is not SoC and Ethernet dependent.
Inspired by the very encouraging results shown in this paper, the authors believe that
this direction could represent a practical solution to the problem of having to recur to expensive
and power consumptive supercomputer every time a scientific and/or engineering numerical task is required.

\bigskip
{\bf{Acknowledgements}}. The authors would like to thank Prof. I.~Dimov for the valuable
conversations. This work has been supported by the project EC AComIn (FP7-REGPOT-2012–2013-1),
as well as by the Bulgarian Science Fund under grant DFNI I02/20.





  \begin{wrapfigure}{l}{25mm} 
    \includegraphics[width=.9in,height=.9in,clip,keepaspectratio]{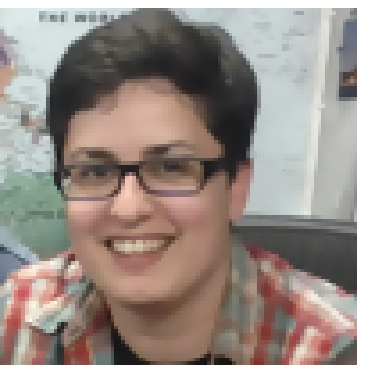}
  \end{wrapfigure}\par
  \textbf{Kristina G. Kapanova} is from the Institute of Information and Communication Technologies at
the Bulgarian Academy of Sciences, where she is currently working to obtain her PhD degree. Her research
interests include numerical methods, artificial neural networks, optimization, system on a chip
development to enable science and engineering research on a smaller scale.\par
\bigskip
  \begin{wrapfigure}{l}{25mm}
    \includegraphics[width=.9in,height=.9in,clip,keepaspectratio]{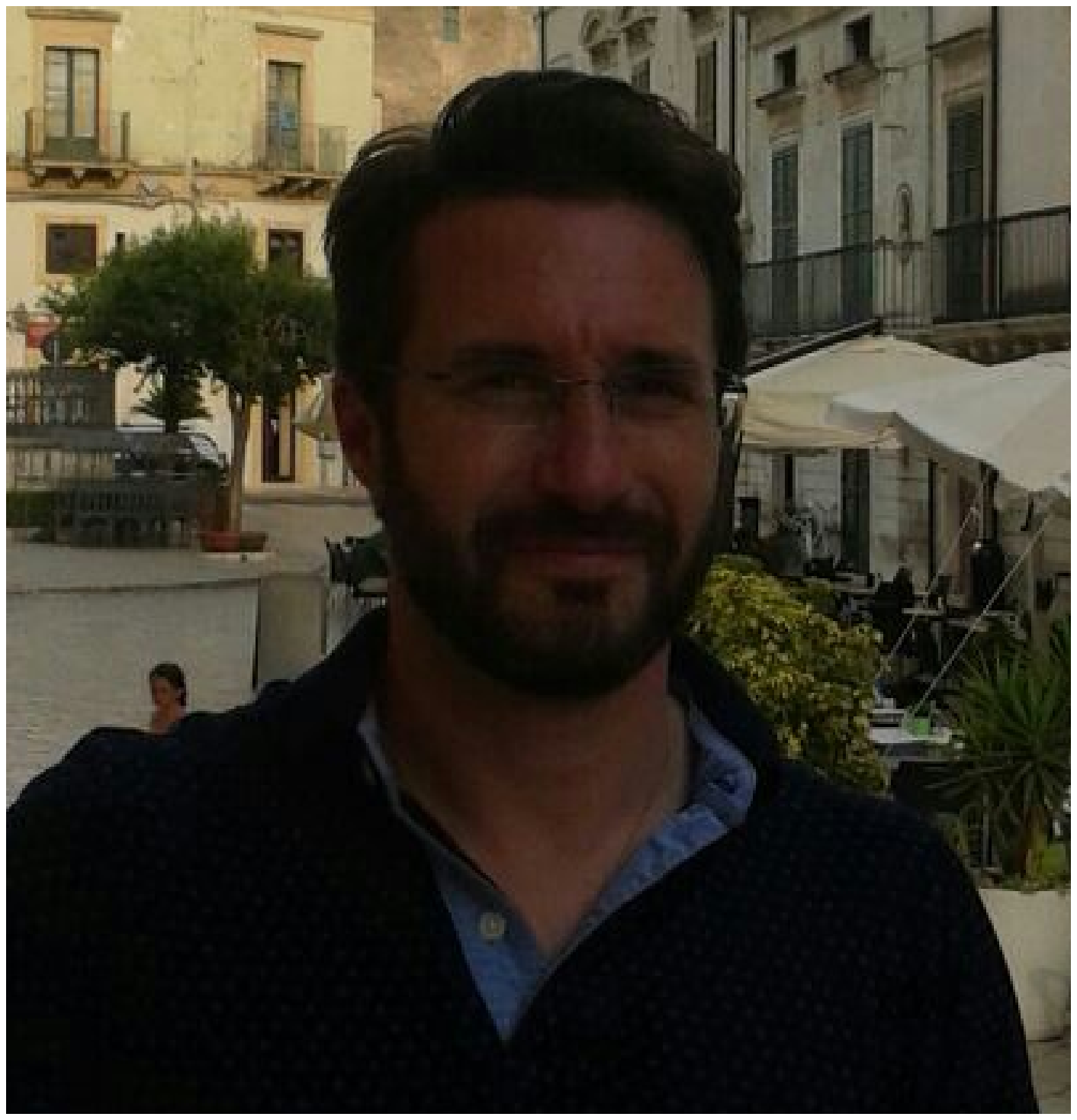}
  \end{wrapfigure}\par
  \textbf{Jean Michel Sellier} is currently an Associate Professor at the Bulgarian Academy of Sciences.
He is also the creator and maintainer of several GNU packages for the simulation of electron transport
in CMOS devices (Archimedes), and the simulation of single- and many-body quantum systems occurring in
the field of quantum computing, spintronics, quantum chemistry and post-CMOS devices (nano-archimedes).\par

\end{document}